Article

# A planet within the debris disk around the pre-main-sequence star AU Microscopii

Peter Plavchan[1✉], Thomas Barclay[2,13], Jonathan Gagné[3], Peter Gao[4], Bryson Cale[1], William Matzko[1], Diana Dragomir[5,6], Sam Quinn[7], Dax Feliz[8], Keivan Stassun[8], Ian J. M. Crossfield[5,9], David A. Berardo[5], David W. Latham[7], Ben Tieu[1], Guillem Anglada-Escudé[10], George Ricker[5], Roland Vanderspek[5], Sara Seager[5], Joshua N. Winn[11], Jon M. Jenkins[12], Stephen Rinehart[13], Akshata Krishnamurthy[5], Scott Dynes[5], John Doty[13], Fred Adams[14], Dennis A. Afanasev[13], Chas Beichman[15,16], Mike Bottom[17], Brendan P. Bowler[18], Carolyn Brinkworth[19], Carolyn J. Brown[20], Andrew Cancino[21], David R. Ciardi[16], Mark Clampin[13], Jake T. Clark[20], Karen Collins[7], Cassy Davison[22], Daniel Foreman-Mackey[23], Elise Furlan[15], Eric J. Gaidos[24], Claire Geneser[25], Frank Giddens[21], Emily Gilbert[26], Ryan Hall[22], Coel Hellier[27], Todd Henry[28], Jonathan Horner[20], Andrew W. Howard[29], Chelsea Huang[5], Joseph Huber[21], Stephen R. Kane[30], Matthew Kenworthy[31], John Kielkopf[32], David Kipping[33], Chris Klenke[21], Ethan Kruse[13], Natasha Latouf[1], Patrick Lowrance[34], Bertrand Mennesson[15], Matthew Mengel[20], Sean M. Mills[29], Tim Morton[35], Norio Narita[36,37,38,39,40], Elisabeth Newton[41], America Nishimoto[21], Jack Okumura[20], Enric Palle[40], Joshua Pepper[42], Elisa V. Quintana[13], Aki Roberge[13], Veronica Roccatagliata[43,44,45], Joshua E. Schlieder[13], Angelle Tanner[25], Johanna Teske[46], C. G. Tinney[47], Andrew Vanderburg[18], Kaspar von Braun[48], Bernie Walp[49], Jason Wang[4,29], Sharon Xuesong Wang[46], Denise Weigand[21], Russel White[22], Robert A. Wittenmyer[20], Duncan J. Wright[20], Allison Youngblood[13], Hui Zhang[50] & Perri Zilberman[51]

AU Microscopii (AU Mic) is the second closest pre-main-sequence star, at a distance of 9.79 parsecs and with an age of 22 million years[1]. AU Mic possesses a relatively rare[2] and spatially resolved[3] edge-on debris disk extending from about 35 to 210 astronomical units from the star[4], and with clumps exhibiting non-Keplerian motion[5–7]. Detection of newly formed planets around such a star is challenged by the presence of spots, plage, flares and other manifestations of magnetic 'activity' on the star[8,9]. Here we report observations of a planet transiting AU Mic. The transiting planet, AU Mic b, has an orbital period of 8.46 days, an orbital distance of 0.07 astronomical units, a radius of 0.4 Jupiter radii, and a mass of less than 0.18 Jupiter masses at $3\sigma$ confidence. Our observations of a planet co-existing with a debris disk offer the opportunity to test the predictions of current models of planet formation and evolution.

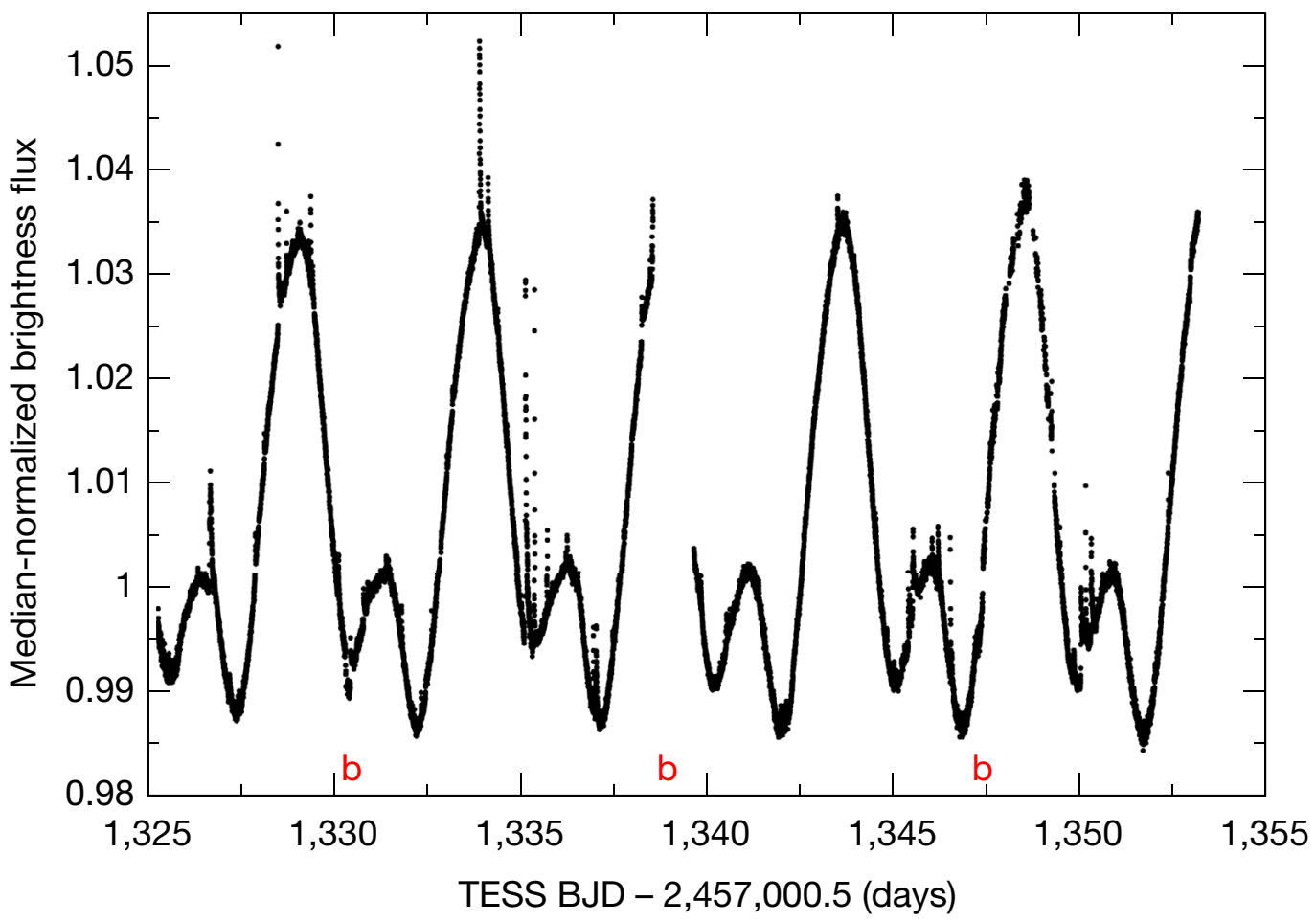

**Fig. 1 | TESS light curve for AU Mic.** Black dots, normalized flux as a function of time, obtained from the MAST archive. Transit ephemerides of AU Mic b are indicated as 'b' in red. The double-humped sinusoidal-like pattern is due to the rotational modulation of starspots, with the 4.863-day rotation period readily apparent. The large, brief vertical streaks of data points deviating upwards from this slower modulation are due to flares. Data with non-zero quality flags indicating the presence of spacecraft-related artefacts, such as momentum dumps (see Fig. 2 legend), are removed. The gap at about 1,339 days corresponds to a gap in the data downlink with Earth during the spacecraft's perigee. A third transit of AU Mic b was missed during this data downlink data gap, and thus the orbital period of AU Mic b is one-half of the period inferred from the two TESS transit events seen. AU Mic exhibited flaring activity with energies ranging from $10^{31.6}$ to $10^{33.7}$ erg in the TESS bandpass over the 27-day light curve (±~60%), with a mean flare amplitude of 0.01 relative flux units. $1\sigma$ measurement uncertainties are smaller than the symbols shown (<1 parts per thousand, p.p.t.).

A list of affiliations appears at the end of the paper.

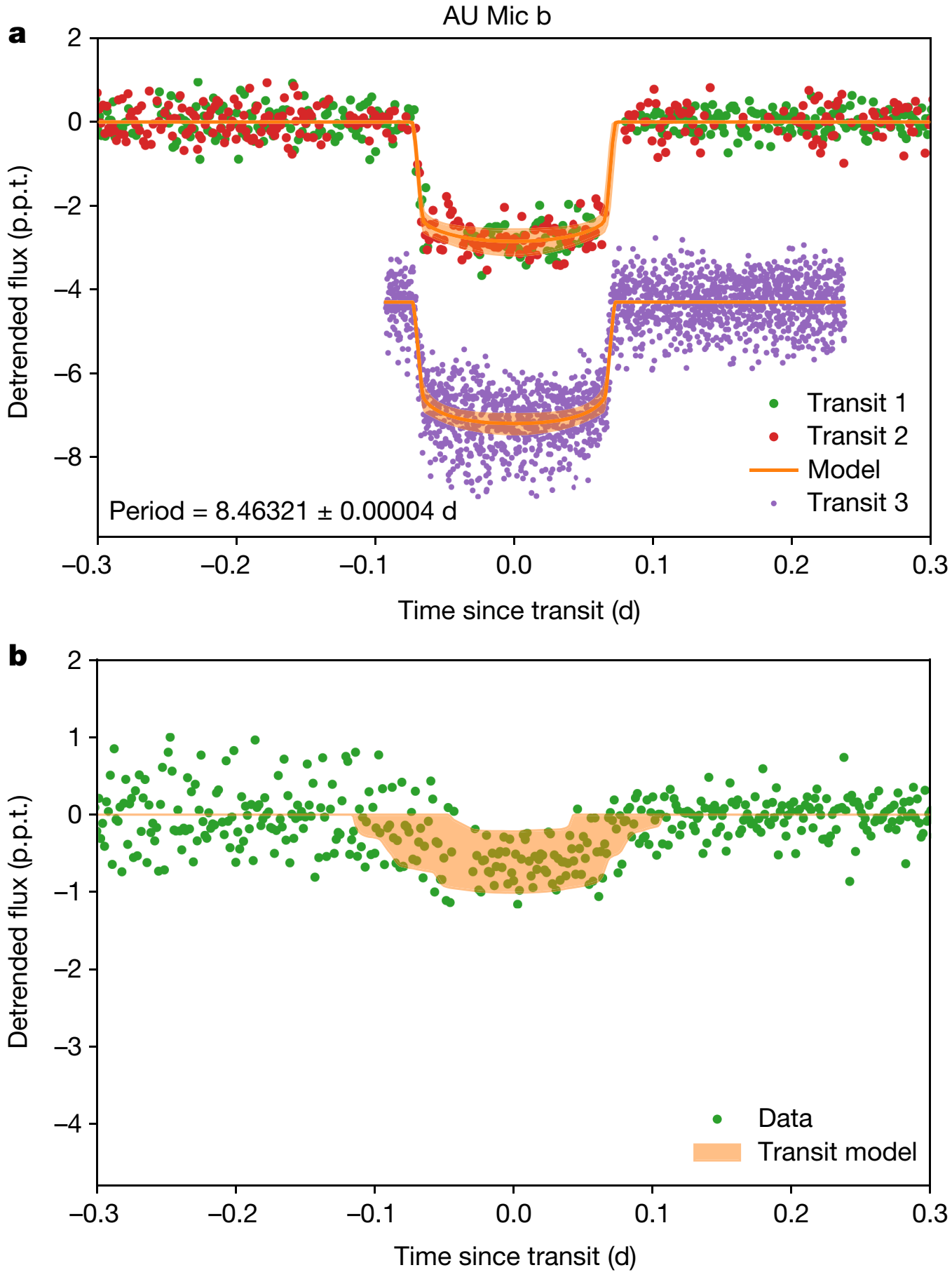

**Fig. 2 | Light curves of the transits of AU Mic b, and a separate, candidate transit event. a**, Data points show light curves from TESS in visible light (green and red filled circles for transits 1 and 2, respectively) and from Spitzer IRAC[11] at 4.5 µm wavelength (purple filled circles for transit 3). The data for transits of AU Mic b are shown with an arbitrary vertical shift applied for clarity; flux units are p.p.t. The transit model (orange curve) includes a photometric model that accounts for the stellar activity modelled with a Gaussian Process (GP), which is subtracted from the data before plotting. The frequent flares from the stellar surface are removed with an iterative sigma-clipping (see Methods). In particular, flares are observed during the egress of both the TESS transits of AU Mic b, and also just after the ingress of the second transit of AU Mic b. The presence of these flares in the light curve particularly affect our precision in measuring the transit duration and thus the mass/density of the host star AU Mic, and consequently the impact parameter and eccentricity of the orbit of AU Mic b. Model uncertainties shown as shaded regions are 1$\sigma$ confidence intervals. The uncertainty in the out-of-transit baseline is about 0.5 p.p.t. but is not shown for clarity. **b**, The AU Mic candidate single transit signal, identified by visual inspection of the TESS light curve. The change in noise before and after the candidate transit signal is due to a 'dump' of angular momentum from the spacecraft reaction wheels which decreased the pointing jitter and improved the photometric precision; data points during the dump are not shown.

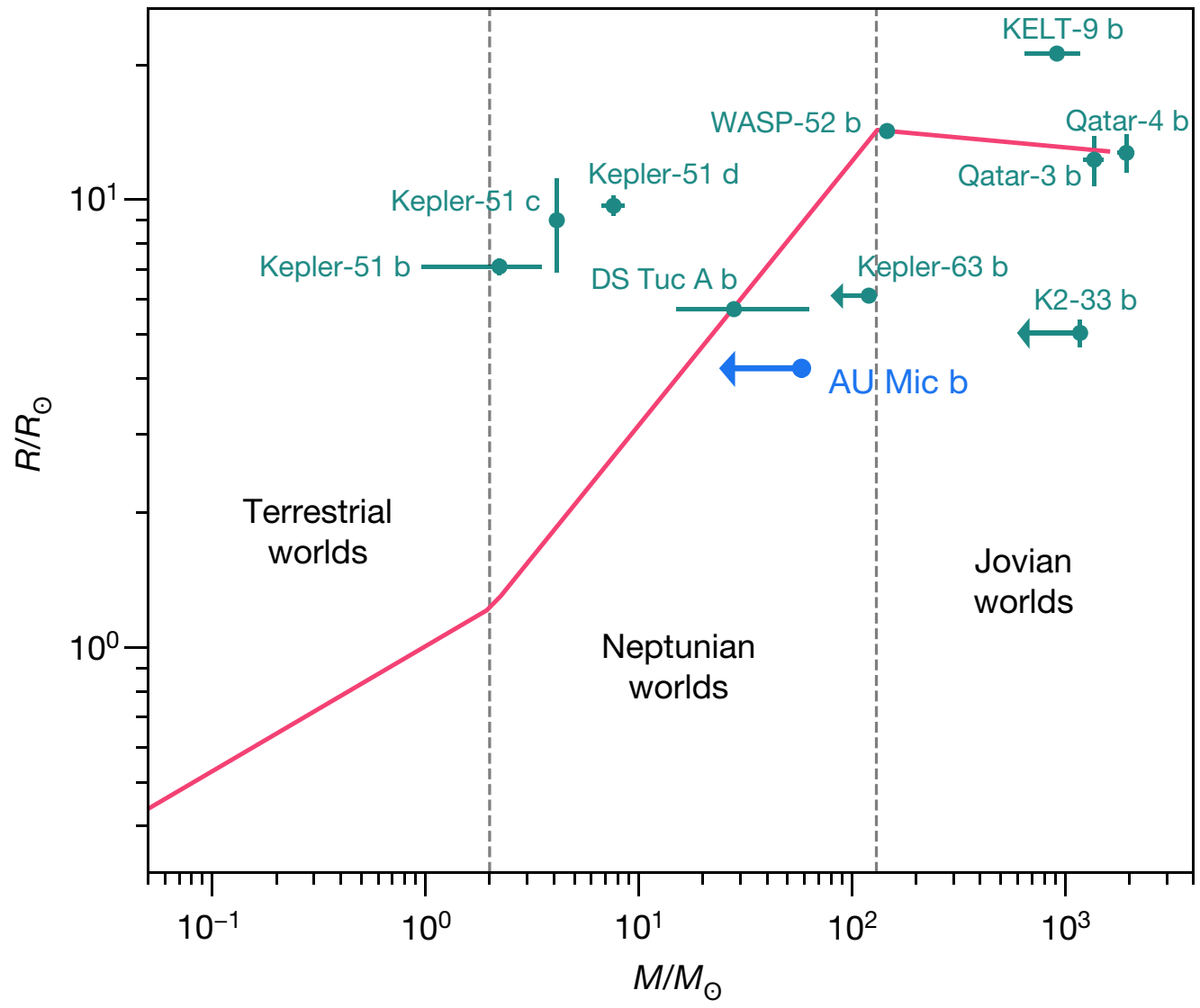

**Fig. 3 | Mass–radius diagram showing AU Mic b in the context of 'mature' exoplanets and known young exoplanets.** Mass *M* and radius *R* are normalized to the values for Earth, respectively $M_\oplus$ and $R_\oplus$. AU Mic b is shown in blue; we compare it to the nominal best-fit mass–radius relationship from known exoplanets orbiting older main-sequence stars[19], shown as a red segmented line (dispersion not shown), and known exoplanets from the NASA Exoplanet Archive with measured masses or mass upper limits, radii, and estimated stellar host ages ≤400 Myr, as follows: DS Tuc A b (mass is estimated from ref. [19] and not measured), Kepler-51 bcd, Kepler 63 b, K2-33 b, Qatar-3 b, Qatar-4 b, KELT-9 and WASP-52 b. By combining the radius measurement from TESS, and the mass upper limit from radial velocities (RVs), we can ascertain an upper limit to the density of AU Mic b to critically inform models for planet formation. Our current upper limit for the mass of AU Mic b cannot rule out a density consistent with Neptune-like planets orbiting older main-sequence stars, but a more precise constraint or measurement in the future may show it to be inflated. Uncertainties shown are 1$\sigma$ for detections, and 3$\sigma$ for mass upper limits.

NASA's Transiting Exoplanet Survey Satellite (TESS) mission[10] was launched on 18 April 2018, and monitored the brightness of AU Mic during the first 27 days of its survey of most of the sky (Fig. 1). Two transits of AU Mic b appear in the TESS photometric light curve. Follow-up observations with the Spitzer Space Telescope[11] confirm the transits of AU Mic b. Our analyses show that this transiting planet has an orbital period of 8.46 days, an orbital distance of 0.07 astronomical units (AU) and a radius of 0.4 Jupiter radii. An additional, shallower candidate transit is observed in the TESS light curve, which suggests the possible existence of additional planets (Fig. 2). Joint radial-velocity (RV) and high-resolution adaptive optics imaging rules out[12] other planets in this system more massive than Jupiter interior to about 20 AU. The 3$\sigma$ upper limit to the velocity reflex motion semi-amplitude for AU Mic b is $K < 28$ m s$^{-1}$ (see Methods), corresponding to an upper limit for the mass of AU Mic b of <0.18 Jupiter masses ($M_{\mathrm{Jupiter}}$) or <3.4 Neptune masses ($M_{\mathrm{Neptune}}$; see Fig. 3, Tables 1 and 2).

The proximity, brightness, age and edge-on geometry of the AU Mic system will permit us to study AU Mic b at an early stage of its dynamical, thermal and atmospheric evolution, as well as any connection between the planet and the residual debris disk. The host star is a red dwarf, one of the most abundant stellar types in our Galaxy. Their diminutive size, mass and luminosity make middle-aged, comparatively inactive M dwarfs favoured targets to search for Earth-size planets in circumstellar habitable zones. Thus AU Mic is an opportunity to study a possible antecedent to these important systems. Moreover, AU Mic, unlike most M dwarfs of a similar age, possesses a debris disk[2], and hence may offer insight into connections between planets and dust disks. This system confirms[13] that gaseous planet formation and any primordial disk migration takes place in less than 20 Myr. The accretion and migration of this (or additional) planets could have left behind the Kuiper-belt-like 'birth ring' of parent body debris that is hypothesized[6] at about 35 AU, while clearing the interior disk of gas and dust. Furthermore, it is possible that any remnant primordial debris in the inner disk near the current locations of the planet could be in the process of being ejected by this planet. Measurement of the spin-orbit obliquity of AU Mic b via the Rossiter–McLaughlin effect (a peak-to-peak amplitude of 40 m s$^{-1}$ is expected) or Doppler tomography would be an important test of migration models since we expect any obliquity in this young system to be unaffected by stellar tides and thus primordial.

**Table 1 | Star parameters**

| Parameter | 68% credible interval | Remarks |
|---|---|---|
| **AU Mic (star)** | | |
| Distance from the Sun | 9.79 ± 0.04 pc | Gaia mission parallax |
| Radius | $(0.75 \pm 0.03)R_{\odot}$ | Directly measured with interferometry[17] |
| Mass | $(0.50 \pm 0.03)M_{\odot}$ | Estimated from spectral type and age[a] |
| $T_{eff}$ | 3,700 ± 100 K | Spectral energy distribution modelling[15] |
| Luminosity | $0.09L_{\odot}$ | Spectral energy distribution modelling[15] |
| Age | 22 ± 3 Myr | Ref. [1] |
| Rotation period | 4.863 ± 0.010 days | RV analysis, TESS light curve, SuperWASP light curve[18] |
| Projected rotational velocity | 8.7 ± 0.2 km $s^{-1}$ | Ref. [12] |
| Linear limb-darkening coefficient (TESS) | $0.21^{+0.20}_{-0.15}$ | TESS light curve |
| Quadratic limb-darkening coefficient (TESS) | $0.0^{+0.18}_{-0.14}$ | TESS light curve |
| Linear limb-darkening coefficient (Spitzer) | $0.17^{0.22}_{-0.12}$ | Spitzer light curve |
| Quadratic limb-darkening coefficient (Spitzer) | $0.15^{+0.27}_{-0.21}$ | Spitzer light curve |
| Visible stellar activity amplitude | $145^{+17}_{-14}$ $ms^{-1}$ | RV analysis |
| Near-infrared stellar activity amplitude | $80^{+16}_{-12}$ $ms^{-1}$ | RV analysis; K band at 2.3 μm |
| Spot decay half-life | 110 ± 30 days | RV analysis |
| GP hyper-parameter 4 | 0.37 ± 0.02 | RV analysis |
| Apparent magnitude | TESS = 6.76 mag | TESS light curve |

[a]Also consistent with independently fitting the two transit events in TESS light curve for AU Mic b.

**Table 2 | Planetary parameters**

| Parameter | 68% credible interval | Remarks |
|---|---|---|
| **AU Mic b** | | |
| Period | 8.46321 ± 0.00004 days | TESS and Spitzer transit light curve analysis |
| Semi-major axis | $0.066^{+0.007}_{-0.006}$AU | TESS and Spitzer transit light curve analysis |
| Velocity semi-amplitude, $K$ | <28 m $s^{-1}$ | RV analysis |
| Mass | $<3.4M_{Neptune}$ $<0.18M_{Jupiter}$ | RV analysis |
| Radius | $(1.08 \pm 0.05)R_{Neptune}$ $(0.375 \pm 0.018)R_{Jupiter}$ | TESS and Spitzer transit light curve |
| Density | <4.4 g $cm^{-3}$ | RV / TESS analysis |
| Time(s) of conjunction | $2,458,330.39153^{+0.00070}_{-0.00068}$BJD[a] | TESS and Spitzer transit light curves |
| Transit duration, $\tau_{14}$ | $3.50^{+0.63}_{-0.59}$h | TESS and Spitzer transit light curves |
| $R_p/R_*$ | 0.0514 ± 0.0013 | TESS and Spitzer transit light curve |
| Impact parameter, $b$ | $0.16^{+0.14}_{-0.11}$ | TESS and Spitzer transit light curve |
| $a/R_*$ | $19.1^{+1.8}_{-1.6}$ | TESS and Spitzer transit light curve |
| Eccentricity | $0.10^{+0.17}_{-0.09}$ | TESS and Spitzer transit light curve[b]. |
| **Candidate transit event** | | |
| Period | 30 ± 6 days | TESS light curve transit duration |
| Radius | $(0.60 \pm 0.17)R_{Neptune}$ = $(0.21 \pm 0.06)R_{Jupiter}$ | TESS transit light curve |
| Time(s) of conjunction | 2,458,342.22 ± 0.03 days | TESS transit light curve |
| $R_p/R_*$ | 0.028 ± 0.006 | TESS transit light curve |
| Impact parameter, $b$ | 0.5 ± 0.3 | TESS transit light curve |
| $a/R_*$ | 40 ± 8 | TESS transit light curve |
| Eccentricity | 0.2 ± 0.2 | TESS transit light curve |

[a]Barycentric Julian Day.
[b]Circular orbit assumed for RV analysis.

AU Mic is a member of the β Pictoris Moving Group; the group's archetype β Pic is a much more massive (about 3.5×), luminous (about 100×) and hotter (approximately 2×) A-type star, also possessing a debris disk. β Pic has a more massive Jovian planet β Pic b observed by direct imaging at a semi-major axis of about 9 AU, with a mass of approximately $(11\pm2)M_{Jupiter}$ determined with astrometry[14]. AU Mic and β Pic are of the same stellar age, but are very different exoplanet host stars. While AU Mic b possibly formed at a distance similar to β Pic b and then migrated inwards to its present location, β Pic b has not substantially migrated inward. These two coeval systems provide an excellent differential comparison for planet formation.

Finally, the combined effect of stellar winds and interior planets have been invoked to explain the high-speed ejection of dust clumps from the system[6,7]. The observed clumps are dynamically decoupled from AU Mic b; the ratio of the semi-major axes (0.06 AU versus >35 AU) is >100, but the clumps could have originated much closer to the star. Dust produced in the debris ring at about 35 AU will spiral inwards primarily as a result of stellar wind drag, which, for AU Mic and a mass loss rate about 1,000 times that of the solar wind[6], is estimated to be 3,700 times stronger than Poynting–Robertson drag[2]. To compare the timescales between collisions of dusty debris and the stellar wind drag force[15], we assume a birth ring fractional width of 10% (3.5 AU), and given AU Mic's infrared flux excess, find that the stellar wind drag and dust collision timescales are roughly equal. Thus, some fraction of the dust grains generated in the birth ring at about 35 AU may spiral inward to the host star under the action of stellar wind drag, instead of being ground down further by dust collisions until blown out of the system by radiation pressure. For 1-μm-sized solid grains of dusty debris, the in-spiral time would be approximately 7,500 years, much shorter than the age of the star. Such dust may have been observed by ALMA[16] at <3 AU, interior to the birth ring at 35 AU. Dust reaching the orbit of an interior planet could be dynamically ejected, depending on the Safronov number: we estimate that of AU Mic b to be 0.07 and thus inefficient at ejecting dust.

There is no other known system that possesses all of these crucial pieces—an M-dwarf star that is young, nearby, still surrounded by a debris disk within which are moving clumps, and orbited by a planet with a direct radius measurement. As such, AU Mic provides a unique laboratory to study and model planet and planetary atmosphere formation and evolution processes in detail.

## Online content

Any methods, additional references, Nature Research reporting summaries, source data, extended data, supplementary information,

acknowledgements, peer review information; details of author contributions and competing interests; and statements of data and code availability are available at https://doi.org/10.1038/s41586-020-2400-z.

[1]Department of Physics and Astronomy, George Mason University, Fairfax, VA, USA. [2]Center for Space Sciences and Technology, University of Maryland Baltimore County (UMBC), Baltimore, MD, USA. [3]Institute for Research on Exoplanets, Département de Physique, Université de Montréal, Montréal, Quebec, Canada. [4]Department of Astronomy, University of California, Berkeley, CA, USA. [5]Massachusetts Institute of Technology, Cambridge, MA, USA. [6]Department of Physics and Astronomy, University of New Mexico, Albuquerque, NM, USA. [7]Harvard-Smithsonian Center for Astrophysics, Cambridge, MA, USA. [8]Department of Physics and Astronomy, Vanderbilt University, Nashville, TN, USA. [9]Department of Physics and Astronomy, University of Kansas, Lawrence, KS, USA. [10]School of Physics and Astronomy, Queen Mary, University of London, London, UK. [11]Department of Astrophysical Sciences, Princeton University, Princeton, NJ, USA. [12]SETI Institute, Mountain View, CA, USA. [13]Exoplanets and Stellar Astrophysics Laboratory, NASA Goddard Space Flight Center, Greenbelt, MD, USA. [14]Department of Astronomy, University of Michigan, Ann Arbor, MI, USA. [15]Jet Propulsion Laboratory, California Institute of Technology, Pasadena, CA, USA. [16]NASA Exoplanet Science Institute, California Institute of Technology, Pasadena, CA, USA. [17]Institute for Astronomy, University of Hawaii at Manoa, Honolulu, HI, USA. [18]Department of Astronomy, University of Texas at Austin, Austin, TX, USA. [19]University Corporation for Atmospheric Research, Boulder, CO, USA. [20]University of Southern Queensland, Centre for Astrophysics, Toowoomba, Queensland, Australia. [21]Department of Physics, Astronomy and Materials Science, Missouri State University, Springfield, MO, USA. [22]Department of Physics and Astronomy, Georgia State University, Atlanta, GA, USA. [23]Center for Computational Astrophysics, Flatiron Institute, New York, NY, USA. [24]Department of Earth Sciences, University of Hawaii at Manoa, Honolulu, HI, USA. [25]Department of Physics and Astronomy, Mississippi State University, Starkville, MS, USA. [26]Department of Astronomy and Astrophysics, University of Chicago, Chicago, IL, USA. [27]Keele University, Keele, Staffordshire, UK. [28]RECONS Institute, Chambersburg, PA, USA. [29]Department of Astronomy, California Institute of Technology, Pasadena, CA, USA. [30]Department of Earth and Planetary Sciences, University of California, Riverside, CA, USA. [31]Leiden Observatory, Leiden University, Leiden, The Netherlands. [32]Department of Physics and Astronomy, University of Louisville, Louisville, KY, USA. [33]Department of Astronomy, Columbia University, New York, NY, USA. [34]IPAC, California Institute of Technology, Pasadena, CA, USA. [35]Astronomy Department, University of Florida, Gainesville, FL, USA. [36]Department of Astronomy, The University of Tokyo, Tokyo, Japan. [37]JST, PRESTO, Tokyo, Japan. [38]Astrobiology Center, NINS, Tokyo, Japan. [39]National Astronomical Observatory of Japan, NINS, Tokyo, Japan. [40]Instituto de Astrofisica de Canarias (IAC), La Laguna, Tenerife, Spain. [41]Department of Physics and Astronomy, Dartmouth College, Hanover, NH, USA. [42]Department of Physics, Lehigh University, Bethlehem, PA, USA. [43]Dipartimento di Fisica "Enrico Fermi", Universita' di Pisa, Pisa, Italy. [44]INAF-Osservatorio Astrofisico di Arcetri, Firenze, Italy. [45]INFN, Sezione di Pisa, Pisa, Italy. [46]Observatories of the Carnegie Institution for Science, Pasadena, CA, USA. [47]Exoplanetary Science at UNSW, School of Physics, UNSW Sydney, New South Wales, Australia. [48]Lowell Observatory, Flagstaff, AZ, USA. [49]NASA Infrared Telescope Facility, Hilo, HI, USA. [50]School of Astronomy and Space Science, Key Laboratory of Ministry of Education, Nanjing University, Nanjing, China. [51]SUNY Stony Brook, Stony Brook, NY, USA.
✉e-mail: pplavcha@gmu.edu

## Methods

### TESS light-curve analysis

AU Mic has long been known as a young star exhibiting flares and brightness variations driven by large starspots on the stellar surface rotating in and out of view[20]. Previous attempts to find transiting planets were not successful owing to this variability and the redness of the star combined with secondary atmospheric extinction effects[21,22], in spite of reasoning that the orbits of any planets could be aligned with AU Mic's edge-on debris disk, and therefore could be more likely to transit than for a random inclination.

TESS observed AU Mic (TIC 441420236) in its first sector (2018 July 25–August 22). The TESS light curve from the 2-min cadence stamp was processed by the Science Processing Operations Center pipeline, a descendant of the Kepler mission pipeline based at the NASA Ames Research Center[23,24]. After visually identifying the transits in the light curve, we independently validate the existence of the transits from the 30-min full-frame image (FFI) data. We also extract light curves with different photometric apertures, and confirm that the transit signal is robust and consistent. No centroid motion is observed during transits, suggesting that it is associated with AU Mic rather than being an instrumental systematic or contamination from scattered background light or a distant star. To validate the transit with ancillary data, we inspect archival sky survey images such as POSS and find no background stars within the TESS pixels that are present at the location of AU Mic with a sufficient brightness ratio so as to mimic the observed transit signals with a background eclipsing binary. Nor do we or others identify any background eclipsing binaries in high-contrast adaptive optics imaging[3] or our high-resolution spectroscopy (see below). The nearest Gaia DR2 source that is capable of producing a false positive if an eclipsing binary (with G-band contrast = 5.7 mag, ignoring TESS-G-band colour terms) is 76 arcsec or 3 TESS pixels from AU Mic. Finally, the interferometric stellar radius determination[17] rules out bound stellar companions.

We perform multiple independent analyses of the TESS light curve to identify and model the transits present, including the TESS mission pipeline planet detection algorithms, ExoFAST v1.0 and v2.0[25,26], and asterodensity profiling[27], which yield consistent results. While ExoFAST does support the simultaneous modelling of light curves and RVs, it does not include components for modelling the stellar activity prevalent for AU Mic in the RVs. Thus, we carry out independent analyses of the light curves and RVs. For the TESS light curve, ExoFAST and astrodensity profiling do not simultaneously model the exoplanet transits and detrending of the photometric variability produced by the rotational modulation of the starspots. Thus to prepare the TESS light curve for these analysis tools, we first fit four sinusoids to the light curve with periods equal to the rotation period, and one-half, one-third and one-quarter thereof. We then apply a 401 data-point running median filter to remove the remaining photometric modulation due to starspots. The flares present in the transit events were not removed for these analyses, primarily affecting the determination of the transit duration of AU Mic b.

### Spitzer light-curve analysis

Owing to the data collection gap in the TESS light curve, Spitzer Director's Discretionary Time (DDT; Program ID no. 14214, 17.3 h time allocation) observations were proposed, awarded and collected in 2019 to rule in or rule out one-half of the orbit period for AU Mic b as seen in the TESS light curve. Three transits were observed with IRAC at 4.5 μm, one of which is presented herein, the others will be presented in a future paper. We first clean up the raw images by sigma-clipping outliers and subtracting off a background estimate from an annulus around the centre of light. We then sum the flux in a circular aperture centred around the centre of light of each frame, and do this for several different aperture radii. We then follow the procedure from ref. [11] and do a pixel level decorrelation (PLD; using 3 × 3 pixels) on each radius, and pick the one that gives the smallest scatter. We adopt a 2.4 pixel radius aperture, binned by a factor of 106.

### Joint TESS and Spitzer photometric analysis

We carry out a custom analysis that simultaneously accounts for the rotational modulation of starspots, the flares and the transit events for both the TESS and Spitzer light curves to evaluate the impact our detrending of the spot rotational modulation and flares has on our analysis of the transit events: this is the analysis we adopt in the main text (Extended Data Fig. 1). We use the TESS pre-search data conditioned light curve created by the TESS pipeline[24,28,29] for this analysis. To remove flares, we create a smoothed version of the light curve by applying a third-order Savitzky–Golay filter with a window of 301 data points, subtracting the smooth light curve, and clipping out data points more deviant than 1.5× the r.m.s. We performed 10 iterations of this clipping, removing the majority of stellar flares. We then used the exoplanet package (https://github.com/dfm/exoplanet) to simultaneously model the stellar variability and transits. Exoplanet uses several other software packages: Starry for the transit model (https://github.com/rodluger/starry) and celerite (https://github.com/dfm/celerite) for the GP, which we use to model stellar variability. Our GP model consists of two terms; a term to capture long-term trends, and a term to capture the periodic modulation of the star's light curve that is caused by spots on the stellar surface. The latter is a mixture of two stochastically-driven, damped harmonic oscillator terms that can be used to model stellar rotation. It has two modes in Fourier space: one at the rotation period of the star and one at half the rotation period. The transit model is parameterized by two stellar limb-darkening parameters, the log of the orbital period, the log of the stellar density, the time of first transit, the log of the planet-to-star radius ratio, the impact parameter of the transit, orbital eccentricity of the planet, and the periastron angle.

We next run a Markov Chain Monte Carlo (MCMC) to fit for the 9 PLD coefficients (the $c_i$s), a slope + quadratic ramp to represent the rotational modulation of the stellar activity still visible for AU Mic in the Spitzer light curve at 4.5 μm, as well as a transit model including two limb-darkening coefficients for a quadratic limb-darkening law (Extended Data Fig. 2). We leave the photometric uncertainty as a free parameter, which we fit for during the MCMC. Prior to the MCMC, we cut out the dip that occurs during the transit, potentially due to a large spot crossing, from Barycentric Modified Julian Date (BMJD) = 58,524.5 to 58,524.53, to make sure we weren't biasing the transit depth. The systematics-corrected light curve is used in our light-curve modelling in the main text.

### Ground-based light-curve analysis

Ref. [21] conducted a dedicated ground-based search for planets transiting AU Mic. One candidate partial transit event ingress was observed (Barycentric Julian Date BJD = 2,453,590.885), with a depth (flux dimming of the star) of ~3%. By itself, this could be attributed to a number of phenomena associated with the star's youth, debris disk, or systematic errors. The photometric precision of this light curve is not sufficient to identify additional transits of AU Mic b or the candidate transit signal from the TESS light curve.

The SuperWASP team monitored AU Mic for seven seasons as part of a larger all-sky survey[22] (Extended Data Fig. 3). We visually inspect the SuperWASP light curve for evidence of any photometry consistent with an ingress or egress from a transiting planet. On several nights, given the ephemeris of AU Mic b, there are photometry data visually similar to an ingress (for example, Julian day (JD) ~2,453,978.40) or an egress (for example, JD ~2,454,232.56). However, the amplitude of the brightness change is comparable to the amplitude of the red (low-frequency) noise in the SuperWASP light curve, and thus these features are probably not real. We do not model or confirm these candidate events, given the stellar activity and relative photometric precision.

The ground-based photometric monitoring[21,22] of AU Mic establishes the long spot lifetimes, which persist for longer than a single observing season as evidenced by the lack of changes in the light curve over many stellar rotations, a defining characteristic of BY Draconis variables. By comparing the TESS, SuperWASP and ref. [21] light curves, it is clear there is spot evolution on a timescale of a few years, as the shape of the phased light curve does differ between the datasets.

### Radial-velocity analysis

Seven RV datasets of AU Mic have been obtained by our team or from the literature and archival data, and a detailed analysis to search for additional planets in the AU Mic system is a subject for future work. In this section, we present the utilization of the higher precision radial velocities from iSHELL, HARPS and HIRES (see below) to rule out higher mass companions, correlations with stellar activity, and confirm the planetary nature of AU Mic b by placing an upper limit on its mass. iSHELL[30] is a near-infrared echelle spectrometer with a resolution of $R = 70,000$ and a simultaneous grasp of a wavelength range of 300 nm at the 3.0-m NASA Infrared Telescope Facility (IRTF); it is equipped with our custom-built methane isotopologue absorption gas cell for wavelength calibration and instrument characterization[31]. The iSHELL data reduction and RV extraction follows the prescription in ref. [31]. We combine our data with archival observations from the visible wavelength HARPS at the ESO La Silla 3.6-m telescope[32], and the visible wavelength HIRES on the 10-m Keck telescope[33] obtained for the California Planet Survey. All HARPS spectra were extracted and calibrated with the standard ESO Data Reduction Software, and RVs were measured using a least-squares template matching technique[34] (Extended Data Figs. 4–6).

AU Mic is very active relative to a main-sequence dwarf, and we find RV peak-to-peak variations in excess of 400 m s$^{-1}$ in the visible range due to the rotational modulation of stellar activity (r.m.s. = 175 m s$^{-1}$ for HIRES and 115 m s$^{-1}$ for HARPS). With iSHELL, the RVs exhibit stellar activity with a smaller but still substantial peak-to-peak amplitude of ~150 m s$^{-1}$ (r.m.s. = 59 m s$^{-1}$). Consequently, no individual RV dataset possesses a statistically significant periodogram signal at the period of planet b. This renders the mass detection of a planet with a velocity semi-amplitude smaller than the activity amplitude challenging[35–38].

We perform an MCMC simulation to model the stellar activity with a Gaussian Process (GP) simultaneously with a circular orbit model for AU Mic b using the regression tool RADVEL[39] (Extended Data Fig. 7). Offsets for the velocity zero point of each RV instrument are modelled. We fix the orbital period and time of transit conjunction (orbital phase) for AU Mic b to the best-fit values constrained by the TESS observations. We assume a velocity semi-amplitude prior with a width of 50% of the best-fit value and positive-definite. Owing to the stellar activity and relatively sparse cadence sampling leading to GP model overfitting, no statistically significant constraints on orbital eccentricity are possible; the eccentricity posterior distributions are unconstrained over the range of eccentricities allowed. Thus, for the sake of brevity we present here only scenarios with fixed circular orbits, although eccentric orbits are considered. Constraining the eccentricity (and periastron angle) of AU Mic b will require a more intensive RV cadence and/or new modelling and mitigation of stellar activity beyond a GP model.

The stellar activity is modelled as a GP with a four 'hyper-parameter' auto-correlation function that accounts for the activity amplitude, the rotation period of the star modulating the starspots, and spot lifetimes treated as an autocorrelation decay[37,40]. From photometric time-series, the spot lifetime for AU Mic is observed to be longer than an observing season. Combined with its known rotation period, this enables us to generate priors on the GP hyper-parameters. We use a Jeffrey's prior on the GP hyper-parameter activity amplitudes bounded between 1 and 400 m s$^{-1}$ for the visible, and 1 and 200 m s$^{-1}$ for the near-infrared, a spot decay lifetime prior that is a Gaussian centred on 110 days with a width of 25 days, a stellar rotation period prior of a Gaussian centred on 4.863 days with a width of 0.005 days, and a Gaussian prior centred on 0.388 with a width of 5% for the fourth hyper-parameter. We assess the dependence of our model comparison on the priors and prior widths used for the planet and GP parameters, which yield qualitatively similar results.

We use the MCMC simulations (Extended Data Fig. 8) to compare statistically favoured models obtained from evaluating the model log-likelihoods, AICc (corrected Akaike information criterion) and BIC (Bayesian Information Criterion) statistics (Extended Data Table 1), and to provide robust characterization of model parameter uncertainties (for example, posterior probability distributions). We derive an upper limit to the velocity reflex motion from AU Mic b of $K < 28.9$ m s$^{-1}$ at $3\sigma$ confidence, corresponding to a mass upper limit of $M_b < 0.18 M_{Jupiter}$ or $<3.4 M_{Neptune}$. We restrict our analysis to estimating an upper limit to the mass of AU Mic b for a number of reasons. First, while our statistical analysis favours the detection of AU Mic b, we do not rule out a non-detection at high statistical confidence. Second, our analysis also relies on the assumption that a GP model is an adequate model for stellar activity. Studies of other starspot-dominated convective M dwarfs[38] suggest this is adequate, but additional future observations and modelling efforts are needed, particularly for stars as active as AU Mic. From Kepler photometric time series of main-sequence stars, we demonstrated[40] that stellar activity should not introduce substantial power in densely sampled (approximately nightly) RV time series at orbital periods longer than the stellar rotation period, as is the case for AU Mic b. However, for more sparsely sampled RV cadences such as ours, stellar activity can introduce apparent periodicities at timescales longer than the stellar rotation period that can persist for several seasons[41]. The long-term magnetic activity evolution of AU Mic on timescales >100 days is also neither constrained nor modelled.

### Wavelength dependence of stellar activity

At near-infrared wavelengths, the expected stellar activity amplitude depends on the effective temperature contrast of the starspots to the photosphere and the effects of Zeeman broadening[35,42]. If the spot temperature contrast is small (for example, a few hundred kelvin), then the RV (and photometric) amplitude due to the rotational modulation of starspots should scale as $1/\lambda$ to first order. This is the case for the Sun[43]. From the HARPS RV r.m.s., one would expect an RV r.m.s. at 2.3 μm of ~50 m s$^{-1}$ if the HARPS RV r.m.s. is entirely ascribable to stellar activity from cool starspots or plages. However, if the spot temperature contrast is large (for example, >1,000 K), one would expect only a marginal (~10%) reduction in RV stellar activity amplitude in the near-infrared. AU Mic lies close to but slightly above the theoretical expectation for cool starspots with small rather than large spot temperature contrast—showing an RV r.m.s. of 59 m s$^{-1}$, a reduction of about two-thirds overall in r.m.s. The modelled GP hyper-parameters for the GP amplitudes show a reduction of about one-half from the visible to the near-infrared.

Ref. [21] obtained multi-band photometry of AU Mic over the course of several rotation periods in their search for transiting exoplanets. Ref. [21] demonstrates that AU Mic exhibits a decreased amplitude of photometric variability as a function of wavelength, again consistent with cool starspots with a relatively small temperature contrast (Extended Data Fig. 9). This is also consistent with multi-band photometry of young pre-main-sequence stars and the Sun[44,45].

### Host star parameters

We compare the mass derived from transit photometry plus Center for High Angular Resolution Astronomy (CHARA) array radius to pre-main-sequence solar-metallicity isochrones of Baraffe et al.[46]. We logarithmically interpolate onto a finer grid, and fit to the absolute J, H and $K_s$ magnitudes (from 2MASS photometry and the Gaia parallax), the radius derived from CHARA[17] and the Gaia parallax, and the effective temperature. The best-fit ($\chi^2 = 20.7$, $\nu = 3$) age and mass are 19 Myr and 0.58$M_\odot$; the uncertainties in age and mass are highly correlated, with a 95.4% confidence interval that spans 9–25 Myr, and (0.38–0.63)$M_\odot$.

### Future work

Additional RVs are necessary to increase the statistical confidence in the determination and recovery of the orbital parameters for AU Mic b and to search for additional planets. In particular, red-sensitive and near-infrared RVs with a nightly monitoring campaign for at least one season are necessary given the relatively large amplitude and timescale of stellar activity, and if possible to search for additional Neptune-mass and smaller planets. Near-simultaneous chromatic RVs, taken at multiple wavelengths across the visible and near-infrared, and/or polarimetric observations may enable a future analysis that more robustly models the stellar activity than can be accomplished with GP and the non-simultaneous multi-wavelength RVs presented here. Simultaneous multi-wavelength RVs could isolate the chromatic stellar activity signal from the achromatic planet signals. Additionally, AU Mic has a $v\sin i$ value of 8.7 km s$^{-1}$, and Zeeman Doppler imaging may enable a mapping of the spot configuration on the stellar surface of AU Mic to monitor long-term activity changes.

Future ground- and space-based photometric monitoring, particularly at red and infrared wavelengths, are needed to further constrain the transit parameters. Observing transit timing variations (TTVs) may be possible for this system to search for additional planets, but the analysis will be complicated by the rotational modulation of the starspots and flares. Flares occur frequently during transit, and since AU Mic b potentially crosses active features on the stellar surface, this renders precise transit depth and duration measurements challenging. Here again, simultaneous multi-wavelength photometry could assist in distinguishing the transit signal from stellar activity. In particular, the Spitzer light curve presented here and planned future observations will provide insights into the spot structure of the surface of AU Mic from spot-crossings by AU Mic b for cross-comparison with the Zeeman Doppler imaging maps.

AU Mic b is also an interesting target to search for signatures of its atmosphere, and for extended hydrogen or helium exospheres, with multiple existing and planned near-term instrumentation on the ground and in space. Given its potentially low density, AU Mic b is one of the most favourable targets to search for planetary atmospheres, even taking into account the upper-limit mass measurement. In particular, since the host star AU Mic is a young active star, it may promote the helium mass loss already detected in other Neptune-size bodies[47,48]. Thus, high-dispersion transmission spectroscopy with visible and near-infrared spectrographs, around the 1,083 nm He I and the Hα line, will measure or constrain atmospheric mass loss rate from this young warm planet.

Since the AU Mic system is young, nearby, possesses a debris disk and is a planet that can be observed in transit, it provides an interesting laboratory to explore several theoretical issues. First, simulations should be carried out of the present and past interactions between the inner planet, the possible inner debris disk at <3 AU (ref. [16]), and the outer debris disk including its clumpy structures[7,49,50]. These interactions depend on the masses of both the outer disk and the inner planet, so that this analysis could provide constraints on their properties; moreover, given the 22 Myr age of the star, these integrations can be carried out over the entire possible age of the stellar system. Second, sensitive searches for trace gas could be carried out for this system. Until a few years ago, the classical definition of a debris disk was the secondary generation of dust. Recently, an increasing number of debris disks have shown gas (today up to 17 sources), including the debris disk orbiting β Pic[51], which is rich in carbon, oxygen and nitrogen, perhaps originating from icy grains rich in CO.

Last, it would be useful to compare the properties of AU Mic b with predictions from planet formation/evolution models. If the mass of AU Mic b is close to our upper limit, the observed radius is close to its expected value for a several Gyr-old planet, whereas the predicted contraction timescale of Neptune-size, gas-rich planets is longer than the age of the system[52,53]. These can be reconciled if the planet is substantially less massive than our upper limit. A better mass limit or determination could place interesting constraints on the entropy of planet formation and early thermal evolution.

## Data availability

In addition to the figure data available, all raw spectroscopic data are available either in the associated observatory archive or upon request from the corresponding author. The TESS light curve is available at the MAST archive, and the SuperWASP light curve is available at the NASA Exoplanet Archive. Source data are provided with this paper.

## Code availability

All code that is not readily available on GitHub is available upon request.

**Acknowledgements** This work was supported by grants to P.P. from NASA (award 16-APROBES16-0020 and support from the Exoplanet Exploration Program) and the National Science Foundation (Astronomy and Astrophysics grant 1716202), the Mount Cuba Astronomical Foundation and George Mason University start-up funds.The NASA Infrared Telescope Facility is operated by the University of Hawaii under contract NNH14CK55B with NASA. Funding for the TESS mission is provided by NASA's Science Mission directorate. Some of the data presented here were obtained at the W. M. Keck Observatory, which is operated as a scientific partnership among the California Institute of Technology, the University of California and NASA. The Observatory was made possible by the generous financial support of the W. M. Keck Foundation. The authors wish to recognize and acknowledge the very significant cultural role and reverence that the summit of Maunakea has always had within the indigenous Hawaiian community. We are most fortunate to have the opportunity to conduct observations from this mountain. This research has made use of the NASA Exoplanet Archive, which is operated by the California Institute of Technology, under contract with NASA under the Exoplanet Exploration Program. Some of the data presented in this paper were obtained from the Mikulski Archive for Space Telescopes (MAST). The Space Telescope Science Institute is operated by the Association of Universities for Research in Astronomy, Inc., under NASA contract NAS5-26555. This research has made use of the services of the ESO Science Archive Facility, based on observations collected at the European Organisation for Astronomical Research in the Southern Hemisphere with the HARPS spectrometer. This work has made use of data from the European Space Agency (ESA) mission Gaia, processed by the Gaia Data Processing and Analysis Consortium (DPAC). Funding for the DPAC has been provided by national institutions, in particular the institutions participating in the Gaia Multilateral Agreement. MINERVA-Australis is supported by Australian Research Council LIEF Grant LE160100001, Discovery Grant DP180100972, Mount Cuba Astronomical Foundation, and institutional partners University of Southern Queensland, MIT, Nanjing University, George Mason University, University of Louisville, University of California Riverside, University of Florida and University of Texas at Austin. This work was partly supported by JSPS KAKENHI grant numbers JP18H01265 and 18H05439, JST PRESTO grant number JPMJPR1775, NSFC grant number 11673011 and MINECO grant ESP2016-80435-C2-2-R. D.D. acknowledges support for this work provided by NASA through Hubble Fellowship grant HST-HF2-51372.001-A awarded by the Space Telescope Science Institute. B.P.B. acknowledges support from National Science Foundation grant AST-1909209. J.W. and P.G. acknowledge support from the Heising-Simons Foundation 51 Pegasi b fellowship.

**Author contributions** P.P.: lead author, principal investigator for CSHELL/iSHELL gas cell and observations, analysis and interpretation. J.G., P.G., B.C., A.T., S.X.W., R.W.: CSHELL/iSHELL data reduction and forward model codes. W.M.: RADVEL analysis. T.B., D.D., S.Q., D.F.-M., E. Gilbert, C. Huang, D.K., E.K., E.V.Q., A.V.: analysis of TESS light curve. K.S., K.C., N.N., E.P., J.P.: follow-up ground-based observations. I.J.M.C., D.A.B., P.L., E.N.: Spitzer light curve. D.F., B.T., C. Hellier: inspection of ground-based light curves. D.W.L.: TRES. G.A.-E.: HARPS. G.R., R.V., S.S., J.N.W., J.M.J.: TESS mission architects. S.R., A.K., S.D., J.T.: TESS mission. F.A., M.C., M.K., A.R., V.R., J.W.: disk physics. D.A.A., J.E.S., A.Y.: flare analysis. C. Beichman, M.B., C. Brinkworth, D.R.C., S.R.K., B.M., S.M.M., K.v.B.: CSHELL/iSHELL instrumentation. B.P.B., C.J.B., J.T.C., J. Horner, J.K., J.O., C.G.T., R.A.W., D.J.W., H.Z.: MINERVA-Australis. A.C., C.D., E.F., C.G., F.G., R.H., T.H., J.H., C.K., N.L., M.M., T.M., A.N., J.T., B.W., D.W., P.Z.: CSHELL/iSHELL observers. E.J.G.: stellar parameters. A.W.H.: Keck HIRES.

**Competing interests** The authors declare no competing interests.

**Additional information**
**Correspondence and requests for materials** should be addressed to P.P.
**Peer review information** *Nature* thanks Suzanne Aigrain and the other, anonymous, reviewer(s) for their contribution to the peer review of this work.
**Reprints and permissions information** is available at http://www.nature.com/reprints.

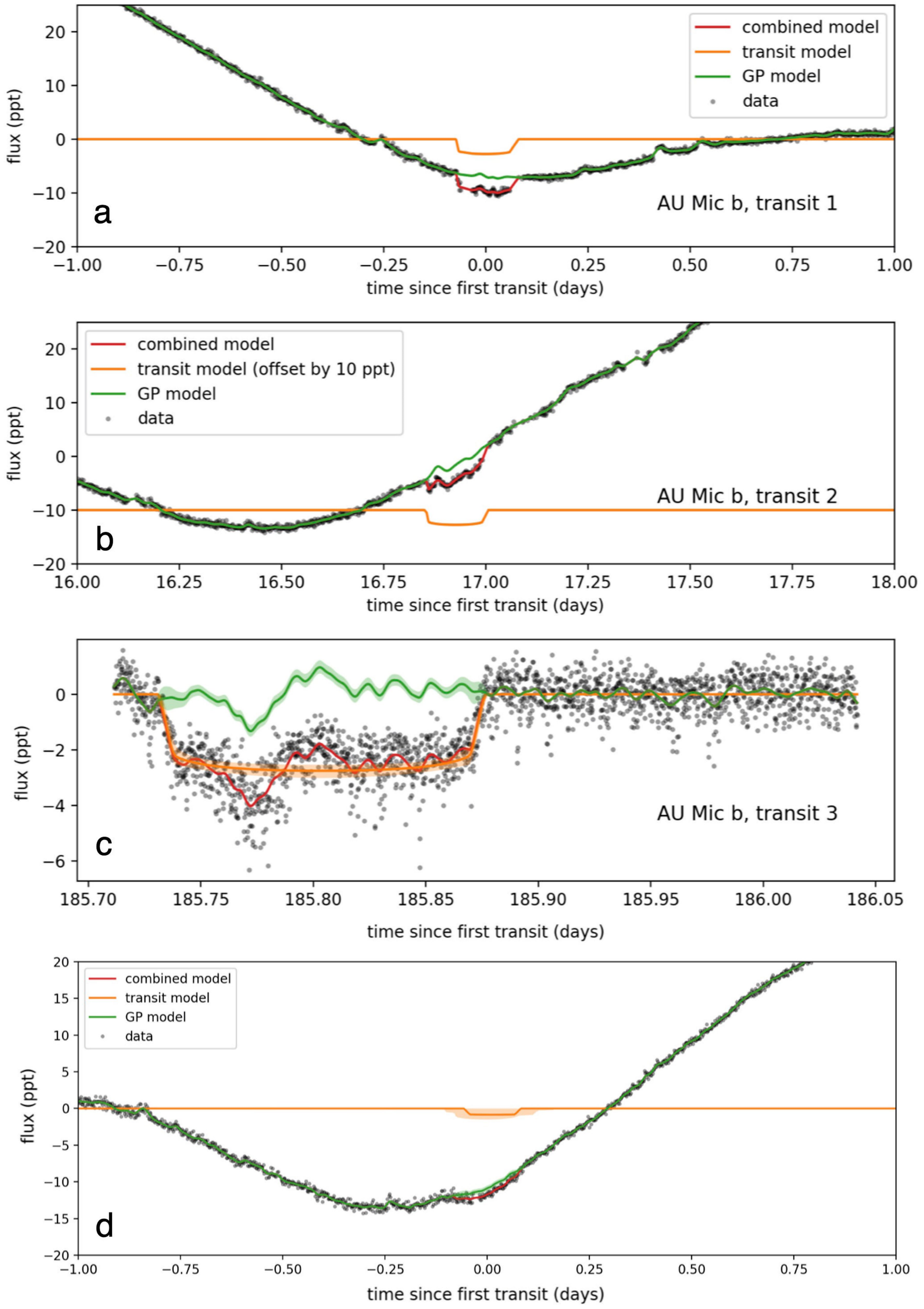

**Extended Data Fig. 1 | TESS and Spitzer light curves for AU Mic centred on four transit events. a**, **b**, Two TESS transits (respectively 1 and 2) for AU Mic b, with the model components plotted as indicated in the key. A flare is present during the egress of the first transit of AU Mic b, and a flare is present just after the ingress during the second transit of AU Mic b. Although this is unfortunate timing, flares of this amplitude are pervasive throughout the TESS light curve for AU Mic, and complicate the recovery of these events from automated transit search algorithms. **c**, The Spitzer transit observation of AU Mic b. The deviations in transit are not instrumental and will be the subject of a future paper, and are likely to be related to the planet crossing large active regions on the stellar surface (key from **a** and **b** applies here). **d**, The ~1 p.p.t. candidate single transit event seen in the TESS light curve. For all panels, 1$\sigma$ measurement uncertainties are suppressed for visual clarity and are <1 p.p.t. 1$\sigma$ model uncertainties in transit are shown as shaded regions.

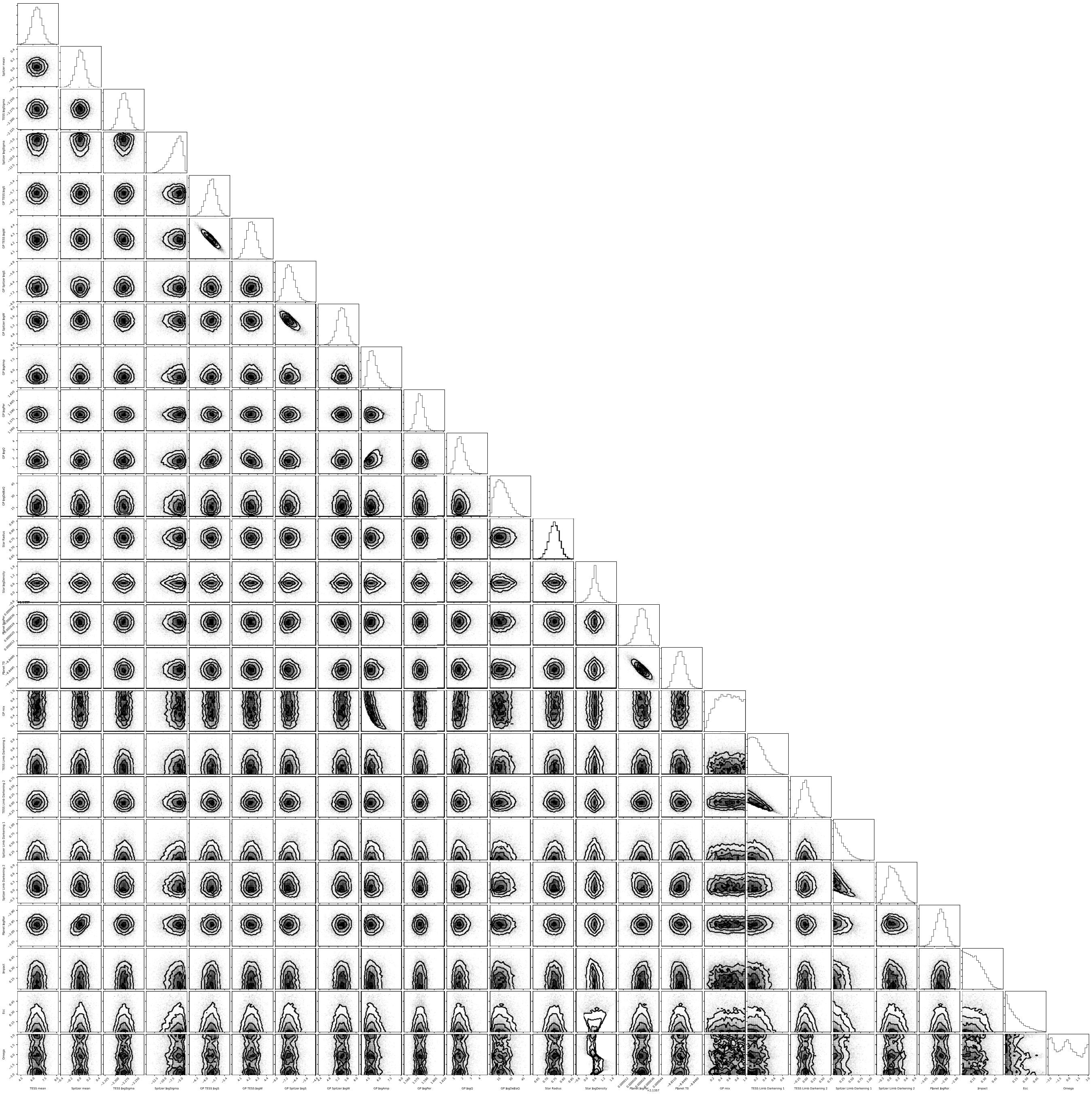

**Extended Data Fig. 2 | MCMC corner plot for custom combined Spitzer and TESS light-curve analysis for AU Mic.** The full set of model parameters are shown, with the posterior probability distributions along the diagonal, the others are the two-dimensional parameter covariance plots.

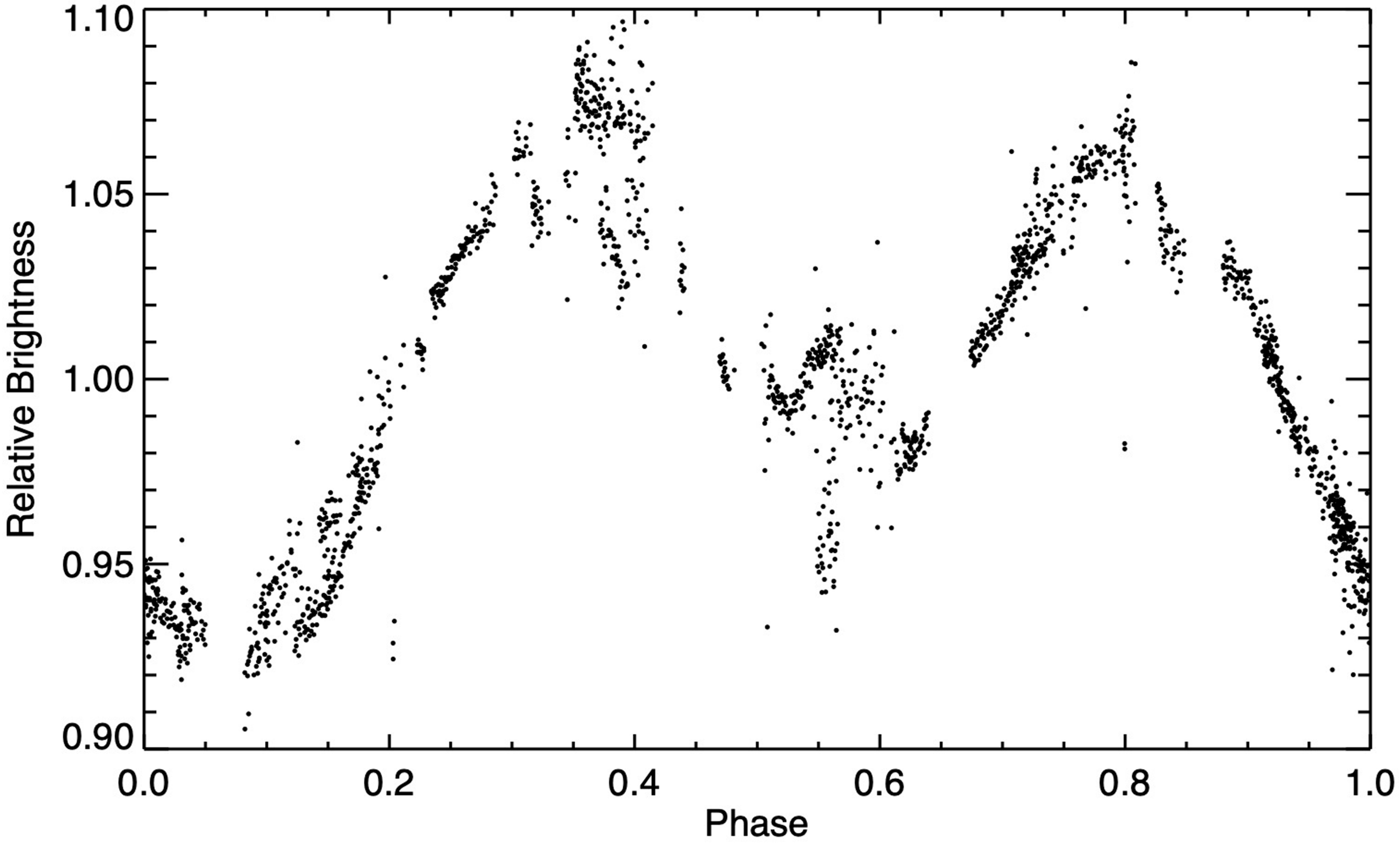

**Extended Data Fig. 3 | One season (July to October 2007) of SuperWASP light curves for AU Mic from the NASA Exoplanet Archive, phase-folded to the rotation period of the star.** Measurements with large photometric uncertainties (>5%) have been excluded from the plot. 1$\sigma$ measurement uncertainties are suppressed for visual clarity and are typically <1% but occasionally up to 5% at phases where there is more apparent vertical scatter in the measurement values themselves.

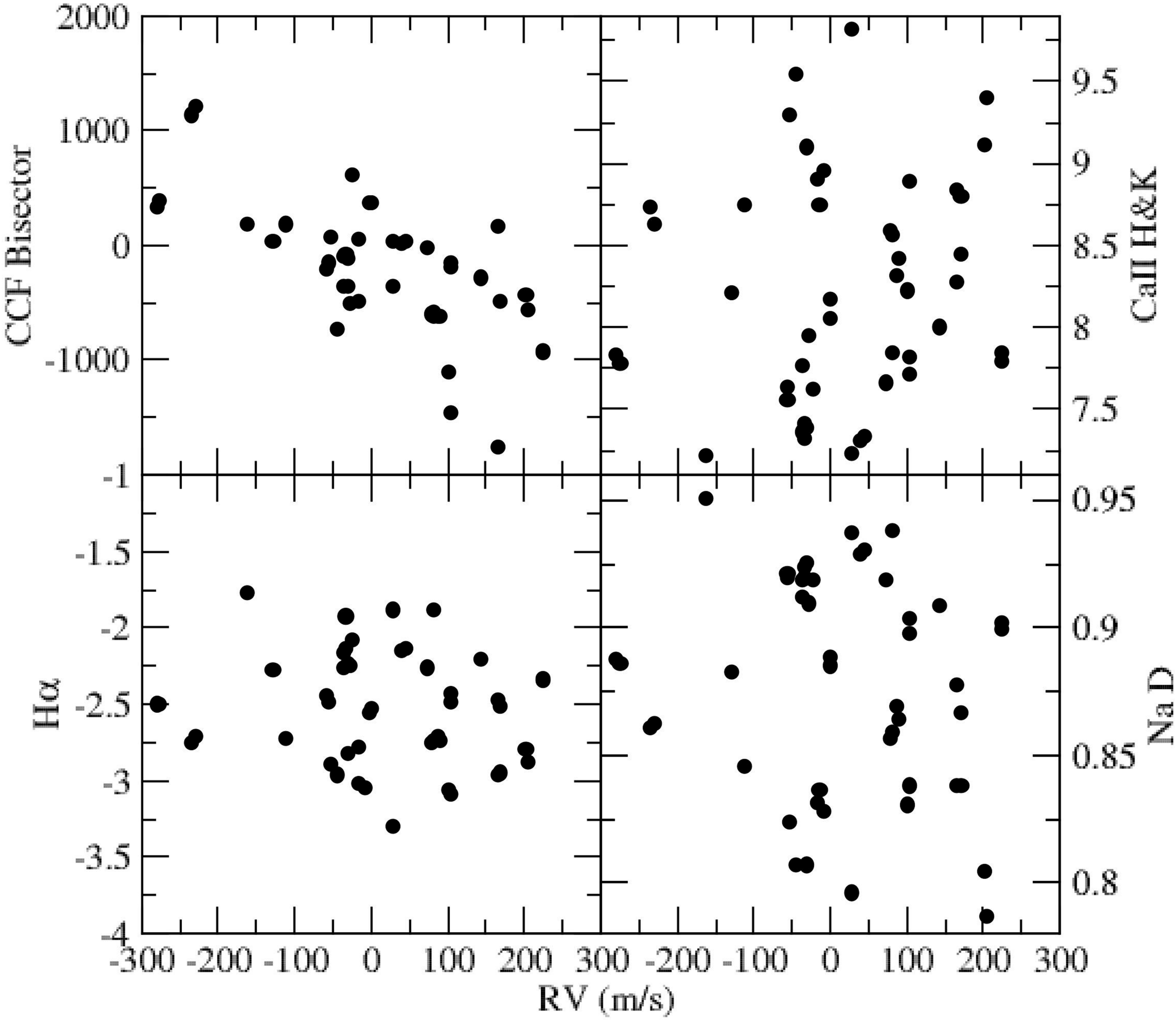

**Extended Data Fig. 4 | Correlation plots of the standard HARPS stellar activity indicators with the RVs.** The bisector values for the cross-correlation function ('CCF bisector'), but not the activity indicators (Hα, Na D, Ca II H and K), show a correlation with the RVs, with substantial remaining scatter. Formal uncertainties are smaller than the plotted symbols.

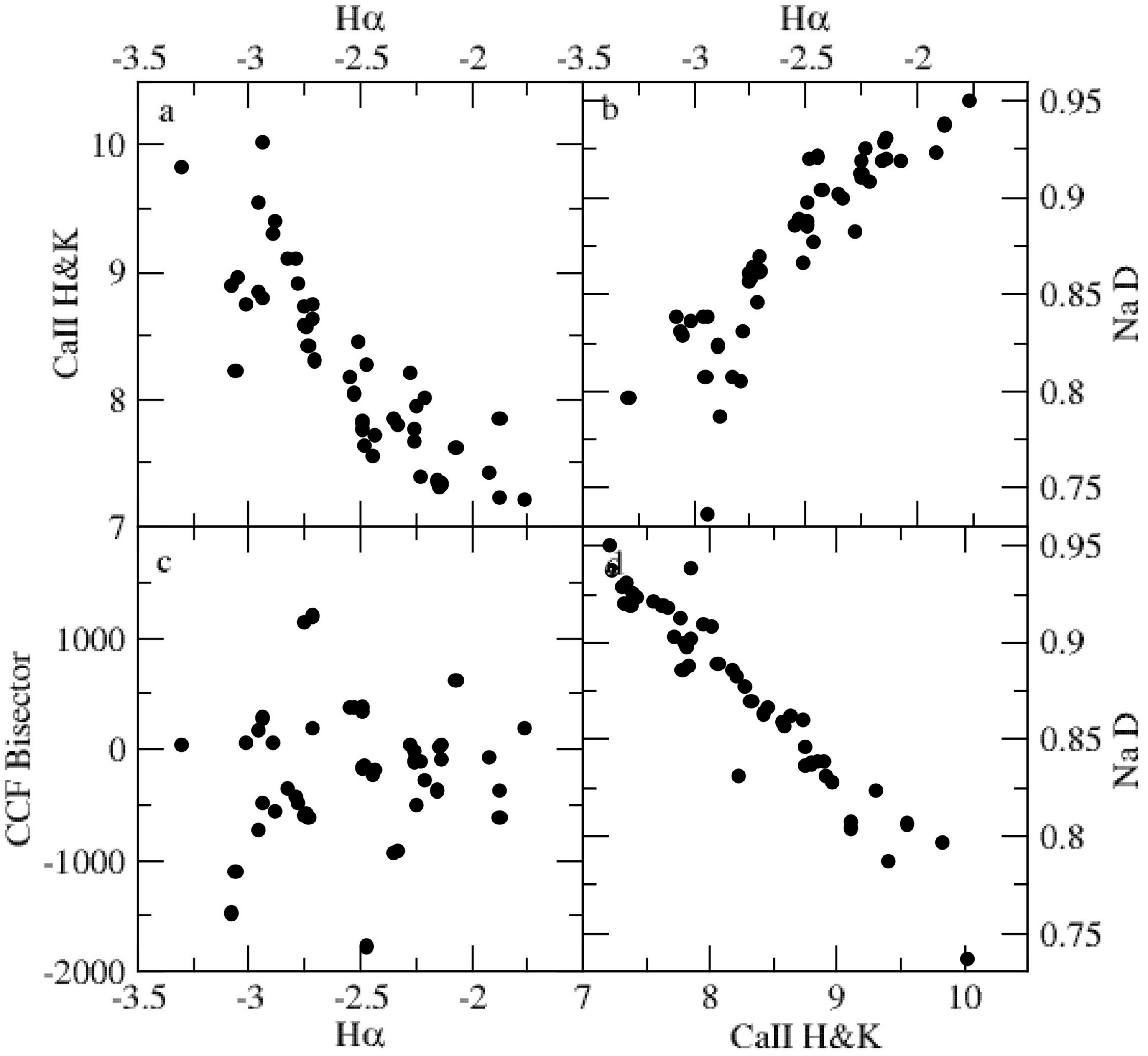

**Extended Data Fig. 5 | Correlation plots of the HARPS activity indicators with each other.** The activity indicators Ca II H and K, Hα, and Na D are strongly correlated with one another, but not with the RVs or with the CCF bisector.

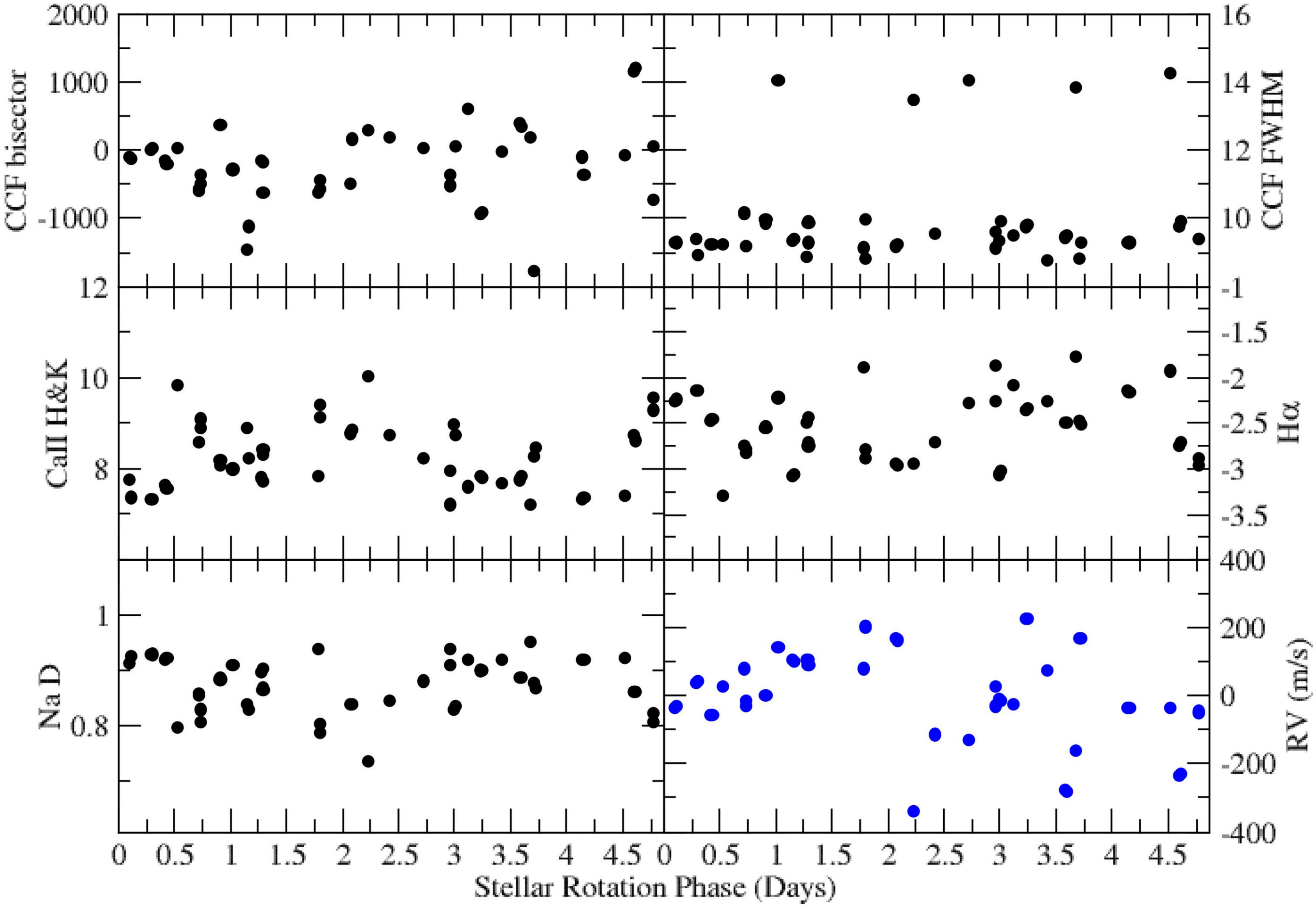

**Extended Data Fig. 6 | The HARPS RVs and standard activity indicators, phase folded to the rotation period of the star.** Blue circles, HARPS RVs; black circles, standard activity indicators. None of the activity indicators show a statistically significant trend with the period of AU Mic b. The Ca and Na activity indicators appear to show (by eye) some cyclic variation with the rotation period of the star. Formal uncertainties are smaller than the plotted symbols.

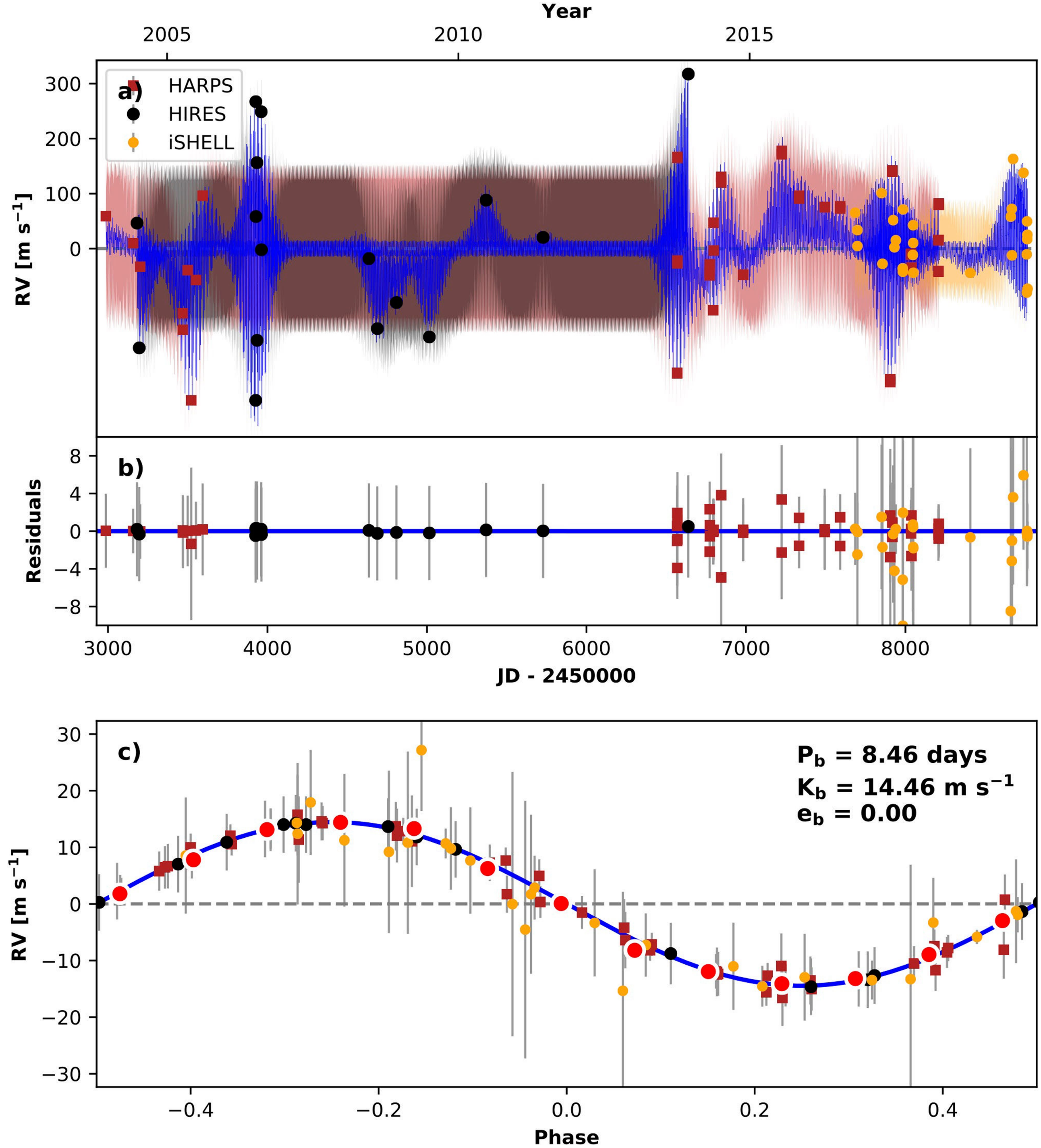

**Extended Data Fig. 7 | RV time-series of AU Mic, with fitting residuals, and phased to the orbital period of AU Mic b.** Shown are data from three spectrometers: iSHELL (yellow circles), HIRES (black circles) and HARPS (red squares). Uncertainties shown are 1$\sigma$ for HARPS and iSHELL. For HIRES, a 5 m s$^{-1}$ minimum 1$\sigma$ uncertainty is adopted, although the formal 1$\sigma$ uncertainties are smaller for all but one epoch at 5.43 m s$^{-1}$. The maximum-likelihood best fit model is overlaid in blue, with shaded regions indicating the 1$\sigma$ model confidence interval, with a separate GP for each dataset indicated with different coloured shaded regions. **b**, Model-subtracted residuals, with the same colours as in **a**. Because our RVs are undersampled with respect to the stellar rotation period[38], the GP best-fit model overfits the AU Mic RV time-series. **c**, RV measurements are phased to the orbital period of AU Mic b, and binned in phase (red circles). The blue curve is a maximum-likelihood best-fit circular orbit model, after subtracting the best fit GP model of stellar activity and the modelled instrument offsets. The plot is labelled with the best-fit orbital period $P_b$, velocity semi-amplitude $K_b$, and the assumed circular orbit ($e_b = 0$).

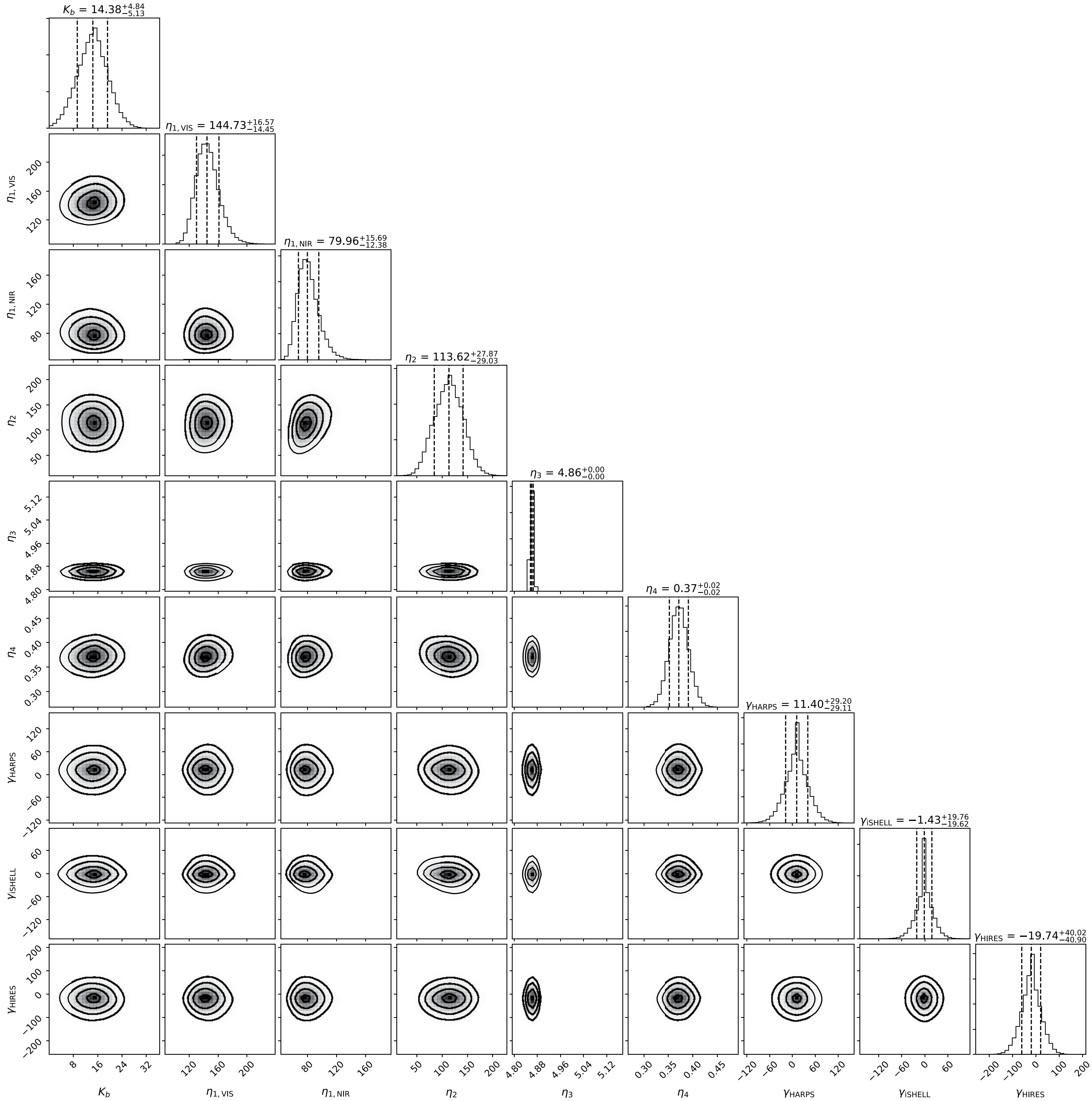

**Extended Data Fig. 8 | RADVEL MCMC corner plot for the model parameters for the iSHELL, HARPS and HIRES RV datasets.** Along the diagonal are the one-dimensional posterior probability distributions for a given model parameter; the others are the two-dimensional parameter covariance plots.

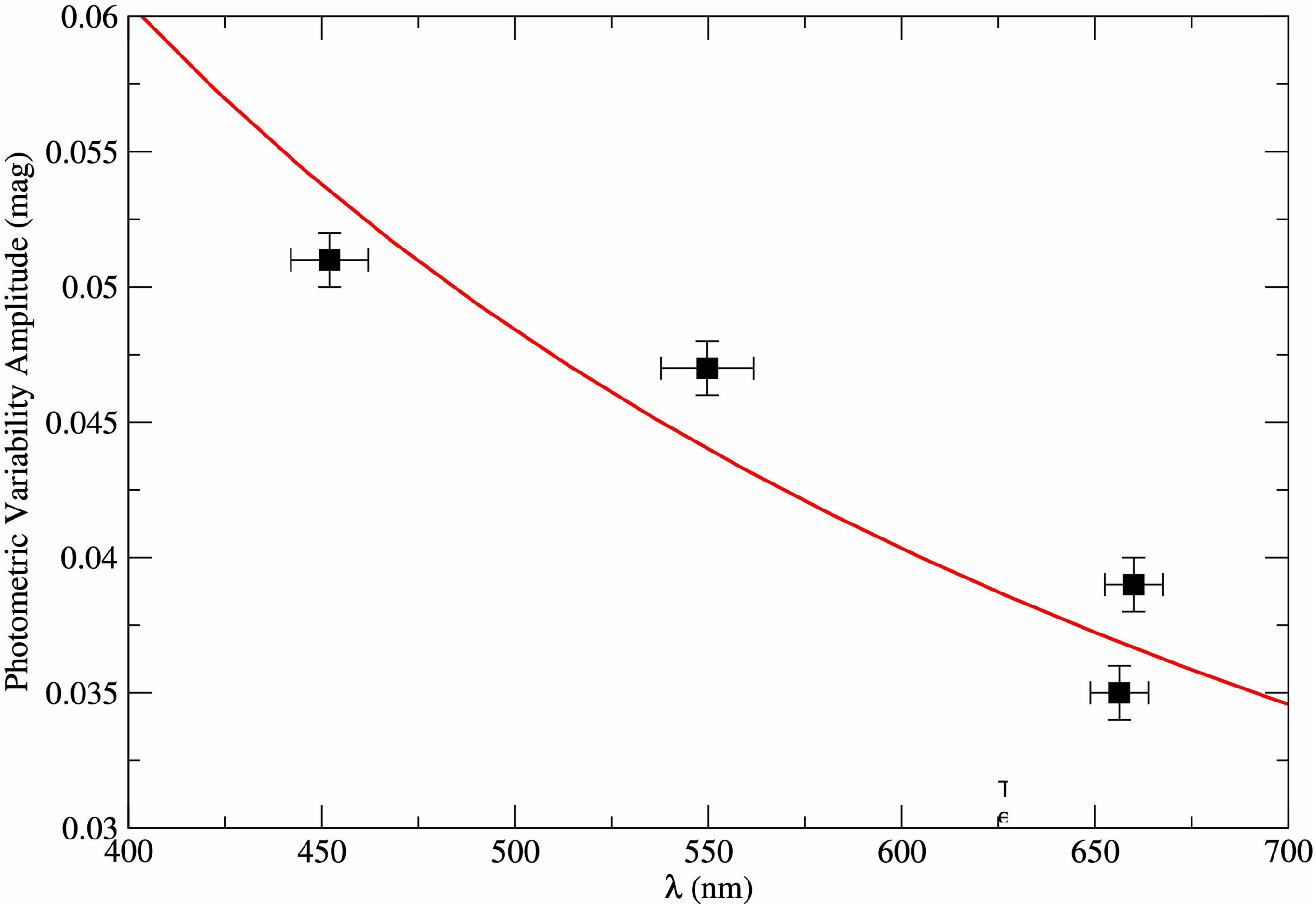

**Extended Data Fig. 9 | Photometric variability amplitudes obtained contemporaneously in four different bandpasses.** The amplitudes (black squares) are from ref. [21]. The horizontal error bars correspond to the effective bandpass widths, and the 1$\sigma$ vertical error bars are set to 1 mmag. A 1/$\lambda$ trend is shown in red, as would be expected for cool starspots with relatively small temperature contrast[35].

**Extended Data Table 1 | Model comparison results**

| AU Mic Model (all include GP & data set offsets) | RV data sets | Free Parameters | Number of RV epochs | Best-fit model rms | log-likelihood | BIC | AICc | Δ AICc between favoured model | AICc qualitative comparison |
|---|---|---|---|---|---|---|---|---|---|
| b | iSHELL, HARPS, HIRES | 9 | 91 | 2.68 | -505.14 | 1050.88 | 1030.50 | 0 | Favoured Model |
| Gaussian Process only | iSHELL, HARPS, HIRES | 8 | 91 | 3.02 | -509.05 | 1054.18 | 1035.85 | 5.35 | Strongly disfavoured |